\documentclass[preprint,showpacs,preprintnumbers,amssymb,superscriptaddress,aps,prd,nofootinbib,11pt]{revtex4-1}
\usepackage{graphicx, graphbox, epsfig, amssymb} 
\usepackage{amsmath, amsfonts}
\usepackage{bm} 
\usepackage{enumitem}
\usepackage[usenames]{color}
\definecolor{navyblue}{rgb}{0.0, 0.0, 0.5}
\usepackage[linktocpage,colorlinks=true,allcolors=navyblue]{hyperref}
\usepackage[caption=false]{subfig}
\usepackage[utf8]{inputenc}
\usepackage{tensor}
\usepackage{soul}

\def\beq{\begin{equation}}
\def\eeq{\end{equation}}

\def\bea{\begin{eqnarray}}
\def\eea{\end{eqnarray}}

\newcommand{\nn}{\nonumber}
\newcommand{\cF}{\mathcal{F}}
\newcommand{\cA}{\mathcal{A}}
\newcommand{\Ftil}{\widetilde{F}}
\newcommand{\bF}{\bm{\mathcal{F}}}
\newcommand{\mbar}{\overline{m}}
\newcommand{\dbar}{\overline{\delta}}
\newcommand{\lnA}{\ln \cA}
\newcommand{\uu}{\mathfrak{u}}
\newcommand{\vv}{\mathfrak{v}}
\newcommand{\ww}{\mathfrak{w}}
\newcommand{\uuj}{\mathfrak{u}_{j}}
\newcommand{\vvj}{\mathfrak{v}_{j}}
\newcommand{\wwj}{\mathfrak{w}_{j}}

\begin{document}\title{\large Higher-order geometrical optics for electromagnetic waves on a curved spacetime}

\author{Sam R. Dolan}\email{s.dolan@sheffield.ac.uk}
\affiliation{Consortium for Fundamental Physics,
School of Mathematics and Statistics,
University of Sheffield, Hicks Building, Hounsfield Road, Sheffield S3 7RH, United Kingdom}

\begin{abstract}
We study the geometrical-optics expansion for circularly-polarized electromagnetic waves propagating on a curved spacetime in general relativity. We show that higher-order corrections to the Faraday and stress-energy tensors may be found via a system of transport equations, in principle. At sub-leading order, the stress-energy tensor possesses terms proportional to the wavelength whose sign depends on the handedness of the circular polarization. Due to such terms, the direction of energy flow is not aligned with the gradient of the eikonal phase, in general, and the wave may carry a transverse stress. 
\end{abstract}

\date{\today}

\maketitle

\section{Introduction\label{sec:intro}}

Our present knowledge of the universe relies on inferences drawn from observations of electromagnetic waves (and, since 2015, gravitational waves \cite{TheLIGOScientific:2017qsa}) that have propagated over astronomical and cosmological distances, across a dynamical curved spacetime. Yet astronomers do not typically analyze Maxwell's equations directly. To account for the gravitational lensing of light, for example, it suffices to employ a (leading-order) geometrical-optics approximation in 4D spacetime \cite{Kristian:1965sz, Seitz:1994xf, Perlick:2004tq}, in which the gradient of the phase is tangent to a light ray. In vacuum, a light ray is a null geodesic of the spacetime. The wave's square amplitude varies in inverse proportion to the transverse area of the beam (as flux is conserved in vacuum) and the polarization is parallel-propagated along the ray (a phenomenon known as gravitational Faraday rotation \cite{Ishihara:1987dv}).

Geometrical optics is a widely-used approximation scheme based around one fundamental assumption: the wavelength (and inverse frequency) is significantly shorter than all other characteristic length (and time) scales \cite{KlineKay, BornWolf:2013}, such as the spacetime curvature scale(s) \cite{Perlick:2004tq}. For the gravitational lensing of electromagnetic radiation, this is typically a good assumption. For example, the Event Horizon Telescope \cite{Johannsen:2016uoh, Akiyama:2017rcc} will seek to image a supermassive black hole of diameter $\sim 10^8$ \text{km} using radiation with a wavelength of $\sim 1$ \text{mm}. On the other hand, gravitational waves have substantially  longer wavelengths (e.g.~$\lambda \sim 10^{7} \text{m}$ for GW150914), as they are generated by bulk motions of compact objects \cite{Abbott:2016bqf}.

The leading-order geometrical-optics approximation will degrade as the wavelength becomes comparable to the space-time curvature scale(s). Formally, Huygen's principle is not valid on a curved spacetime, due back-scattering, and the retarded Green's function has extended support within the lightcone. Nevertheless, wherever there is a moderate separation of scales, one would expect that geometrical-optics would remain a useful guide, and that more accurate results could be obtained by including higher-order corrections in the ratio of scales. Additionally, the structure of the higher-order corrections may provide insight into wave-optical phenomena that are not present in the ray-optics limit. 

Higher-order corrections in geometrical-optics expansions were studied in a pioneering 1976 work by Anile \cite{Anile:1976gq}, building upon the earlier ideas of Ehlers \cite{Ehlers:1967}. By using the spinor and Newman-Penrose formalism, Anile found that ``correct to the first-order, the wave has energy flows in directions orthogonal to the wave's propagation vector, as well as anisotropic stresses.'' This conclusion is perhaps under-appreciated, and may yet find relevance in the era of long-wavelength gravitational wave astronomy. 

In this paper we extend geometrical optics to sub-leading order, in the (4D) spacetime setting, for circularly-polarized waves. Using a tensor formulation, we obtain an expression for the stress-energy tensor $T_{ab}$ that includes the leading-order effect of wave helicity (i.e.~the handedness of circular polarization). Like Anile \cite{Anile:1976gq}, we find a stress-energy that deviates from that of a null fluid at sub-leading order, allowing the circularly-polarized wave to (i) carry transverse stresses due to shearing of the null congruence, and (ii) propagate energy in a direction that is misaligned with the gradient of the eikonal phase. 

A key motivation for this work is the recent interest in a spin-helicity effect: a coupling between the frame-dragging of spacetime outside a rotating body, and the helicity (handedness) of a circularly-polarized wave of finite wavelength \cite{Mashhoon:1974cq}. It is known, for example, that a Kerr black hole can distinguish and separate waves of opposite helicity \cite{Dolan:2008kf, Frolov:2011mh, Frolov:2012zn, Yoo:2012vv}. Similar effects for rotating bodies have also been studied \cite{Barbieri:2004vne, Barbieri:2005kp}. It was suggested in Ref.~\cite{Frolov:2011mh} that this is due to an \emph{optical Magnus effect} which is dual to gravitational Faraday rotation. 

A recent study \cite{Leite:2017zyb} of the absorption of planar electromagnetic waves impinging upon a Kerr black hole, in a direction parallel to the symmetry axis, noted that a circularly-polarized wave with the opposite handedness to the black hole's spin is absorbed to a greater degree than the co-rotating polarization. The difference in absorption cross sections due to helicity was shown to scale in proportion to the wavelength $\lambda$, as $\lambda \rightarrow 0$. This is an example of an effect that is \emph{not} present in the leading-order geometrical-optics limit, but which should be captured by geometrical optics at sub-leading order. 

\begin{figure}
\begin{minipage}[c]{13cm}
 \begin{center}
  \hspace{-5cm}
  \includegraphics[align=c,width=8cm]{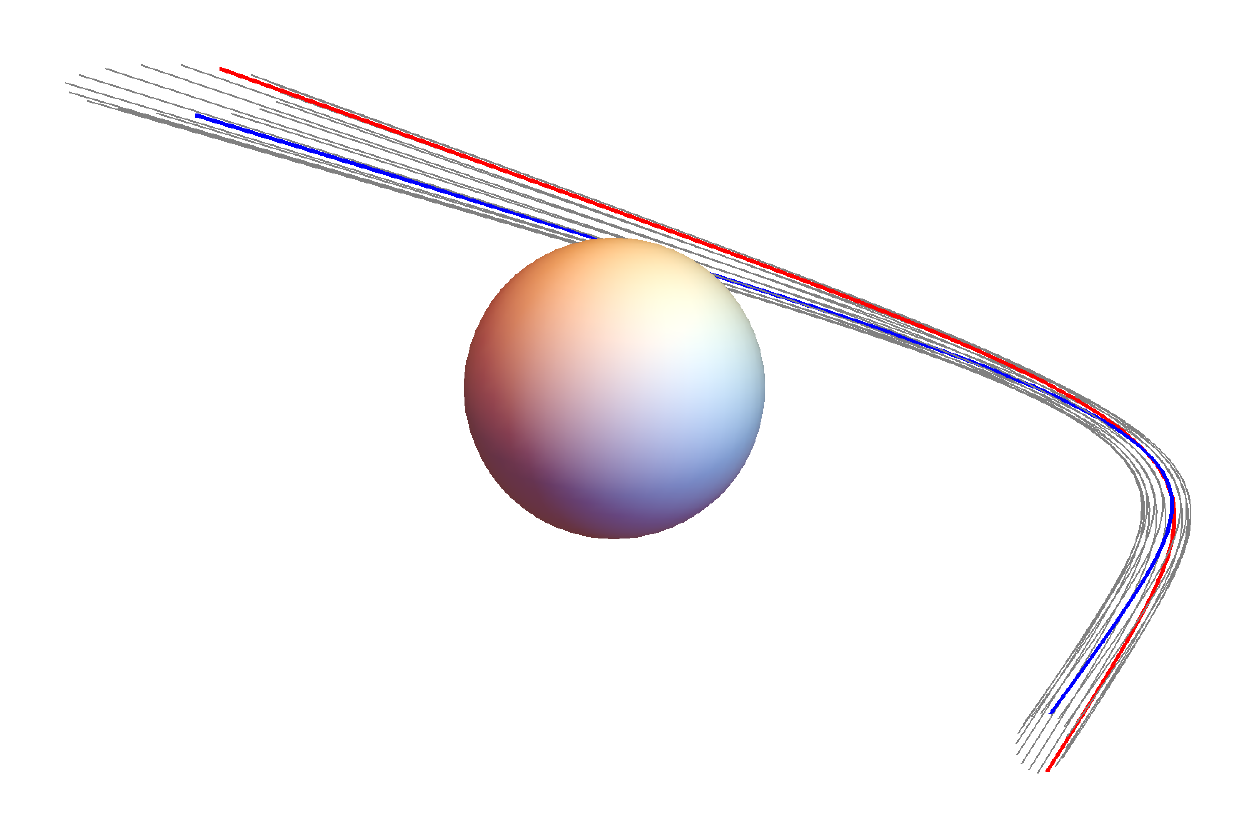}  \\
    \hspace{-5cm}
  \includegraphics[align=c,width=6.5cm]{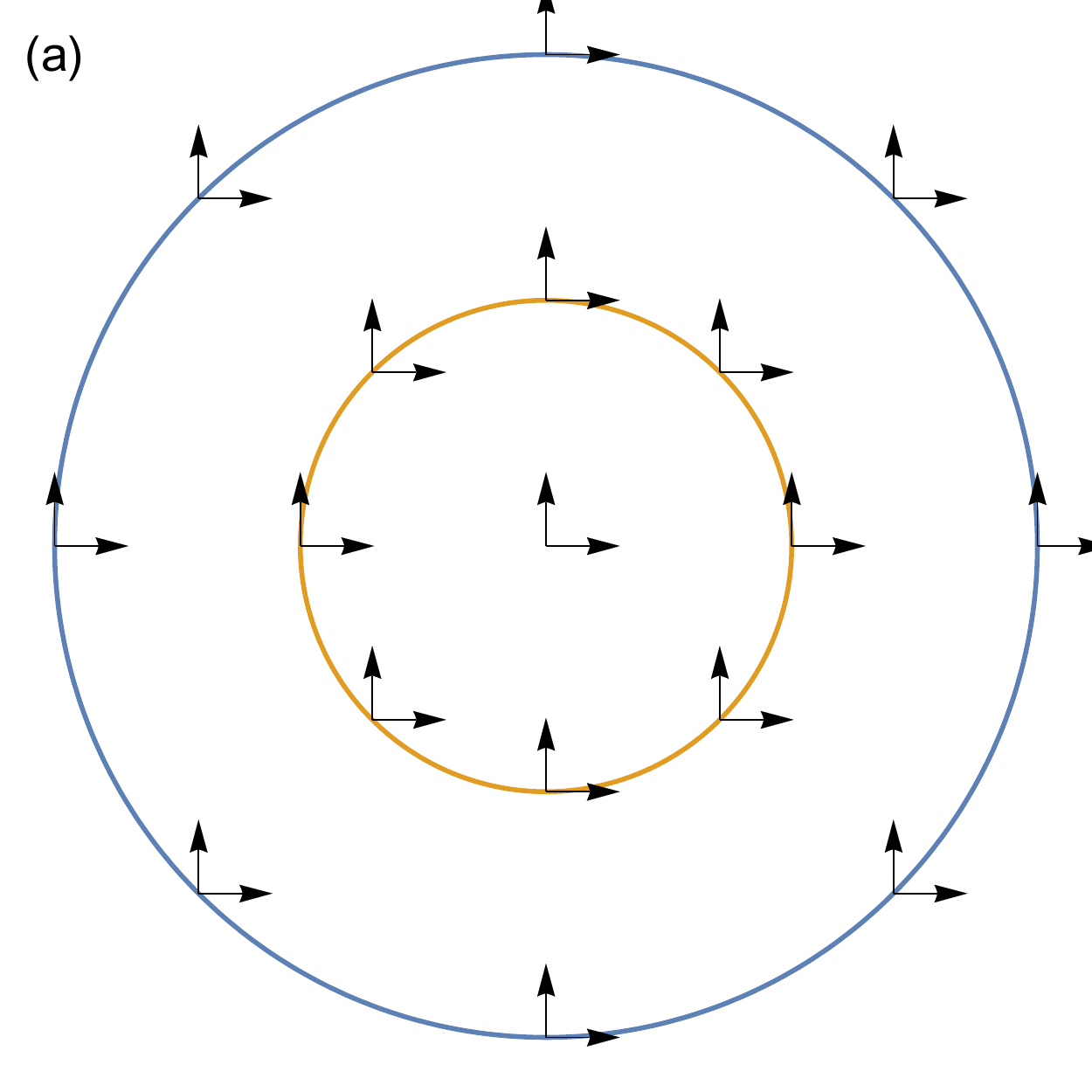}  \; \quad \;
  \includegraphics[align=c,width=5.2cm]{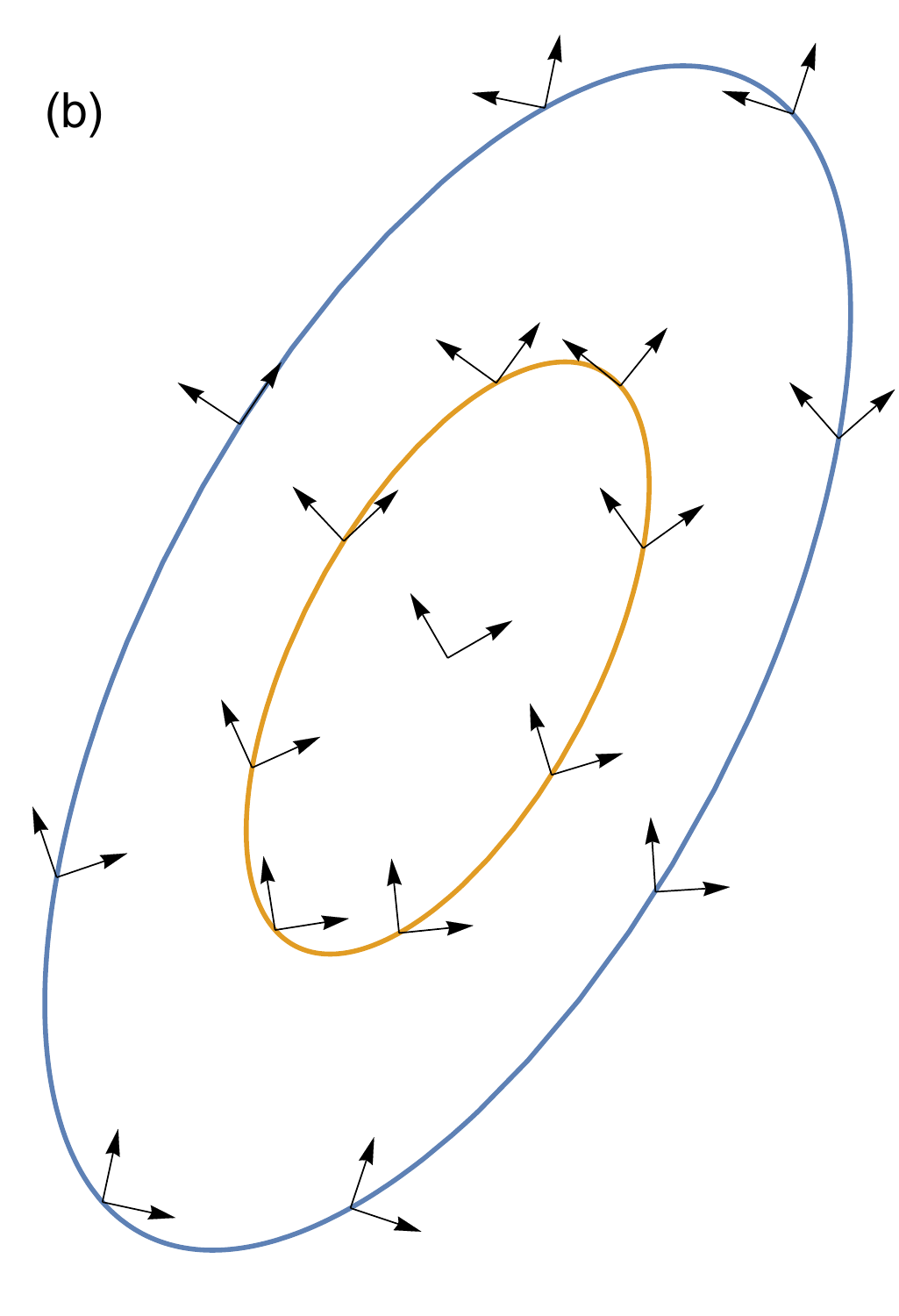}
 \end{center}
\end{minipage}
 \caption{Upper: Showing a congruence of rays deflected in the gravitational field of a compact object. Lower: (a) Circular cross section of a bundle of rays with a basis $m^a = \frac{1}{\sqrt{2}} \left(e_1^a + i e_2^b\right)$ that is `straight': $\mbar^b \nabla_b m^a = 0$.  (b) Elliptical cross section of the same bundle after propagating through a gravitational field.  The parallel-propagated basis $m^a$ has undergone differential precession such that $\mbar^b \nabla_b m^a \neq 0$.}
 \label{fig:differential-precession}
\end{figure}

Here we highlight a mechanism through which a spin-helicity effect could arise, namely, through the differential precession induced in a basis that is parallel-propagated along a null geodesic congruence. The idea is illustrated in Fig.~\ref{fig:differential-precession}, which shows the crosssection of a narrow beam, (a) before and (b) after passing through a gravitational field. The initially-circular crosssection is distorted into an ellipse by geodesic deviation. A set of basis vectors in the crosssection (shown as arrows in Fig.~\ref{fig:differential-precession}) becomes twisted as they are dragged along the rays of the beam. Consequently, the gravitational Faraday rotation angle varies across the crosssection. A local observer would interpret a spatially-varying Faraday rotation as an additional phase that varies across the beam. This additional phase would lead to a correction in the wave's apparent propagation direction. This argument is developed heuristically in Sec.~\ref{subsec:diffprecession}, and put on a firmer footing through the results of Sec.~\ref{sec:higher-order}. 

This paper is organised as follows. Sec.~\ref{sec:foundations} comprises review material:  Sec.~\ref{subsec:maxwell} is on the Faraday tensor, Maxwell's equations, the vector potential, wave equations, and the stress energy tensor; Sec.~\ref{subsec:leading} is on the geometrical-optics approximation at leading order, covering the ansatz for the Faraday tensor, the expansion method, and the resulting hierarchical system of equations; and Sec.~\ref{subsec:congruence} is on the self-dual bivector basis, geodesic deviation, Sachs' equations and the optical scalars.  Sec.~\ref{subsec:diffprecession} concerns the modification to the leading-order phase that arises from differential precession across a null geodesic congruence. Sec.~\ref{sec:higher-order} presents the method for calculating higher-order corrections, in both the tensor formalism (\ref{subsec:subleading}) and the Newman-Penrose formalism (\ref{subsec:NP}). The paper concludes with a discussion of the key results in Sec.~\ref{sec:discussion}. Auxiliary results are presented in Appendix \ref{appendix:NP} and \ref{appendix:Uder}. 

\emph{Conventions:} Here $g_{ab}$ is a metric with signature $-+++$. Units are such that the gravitational constant $G$ and the speed of light $c$ are equal to $1$. Indices are lowered (raised) with the metric (inverse metric), i.e.~$u_a = g_{ab} u^b$ ($u^a = g^{ab} u_b$). Einstein summation convention is assumed. The metric determinant is denoted $g = \text{det}\,g_{ab}$. The letters $a, b, c, \ldots$ are used to denote \emph{spacetime} indices running from $0$ (the temporal component) to $3$, whereas letters $i,j,k,\ldots$ denote \emph{spatial} indices running from $1$ to $3$. The Levi-Civita tensor is $\varepsilon_{abcd} \equiv \sqrt{-g} [a b c d]$, with $[a b c d]$ the fully anti-symmetric Levi-Civita symbol such that $[0 1 2 3] = +1$. The covariant derivative of $X_b$ is denoted by $\nabla_a X_b$ or equivalently $X_{b;a}$, and the partial derivative by $\partial_a X_b$ or $X_{b,a}$. The symmetrization (anti-symmetrization) of indices is indicated with round (square) brackets, e.g.~$X_{(ab)} = \frac{1}{2} (X_{ab} + X_{ba})$ and $X_{[ab]} = \frac{1}{2} (X_{ab} - X_{ba})$. $\{k^a, n^a, m^a, \mbar^a\}$ denote the legs of a (complex) null tetrad. Complex conjugation is denoted with an over-line, or alternatively, with an asterisk: $\mbar^a = m^{a\ast}$. 

\section{Foundations\label{sec:foundations}}

\subsection{Maxwell's equations in spacetime\label{subsec:maxwell}}

\subsubsection{The Faraday tensor}
The fundamental object in electromagnetism is the Faraday tensor $F_{a b}$, a tensor field pervading spacetime which is anti-symmetric in its indices, $F_{b a} = - F_{a b}$ (i.e.~a two-form field). The electric and magnetic fields at a point in spacetime depend on the choice of Lorentz frame. An observer with (unit) tangent vector $u^a$ and (orthonormal) spatial frame $e^a_i$ `sees' an electric field $E_i = F_{a b} e_i^a u^b$ and a magnetic field $B_i = \Ftil_{a b} e_i^a u^b$. Here $\Ftil_{a b}$ is the Hodge dual \cite{Stephani:2003tm} of the Faraday tensor, defined by
\beq
\Ftil_{a b} \equiv \frac{1}{2} \varepsilon_{abcd} F^{cd} ,
\eeq
where $\varepsilon_{abcd}$ is the Levi-Civita tensor. (It follows that $\widetilde{X}_{ab}^{\sim} = -X_{ab}$ for any two-form $X_{ab}$.) 

It is convenient to introduce a complex version of the Faraday tensor,
\beq
\cF_{ab} \equiv F_{ab} + i \widetilde{F}_{ab}.
\eeq
The complex tensor $\cF_{ab}$ is \emph{self-dual}, by virtue of the property $\widetilde{\cF}_{ab} = -i \cF_{ab}$. It follows from its definition that $\cF^\ast_{ab} \cF^{ab} = 0$, where ${}^{\ast}$ denotes complex conjugation.
We may also introduce a complex three-vector $\bF$ with components $\cF_i \equiv \cF_{ab} e_i^a u^b$, whose real and imaginary parts yield the (observer-dependent) electric and magnetic fields, 
$\bF = \mathbf{E} - i \mathbf{B}$.

Under local Lorentz transformations (changes of observer frame), the components of the complex three-vector $\mathbf{\bF}$ transform as follows \cite{Synge:1964}: $\cF_i \rightarrow \cF^\prime_i = O_{ij} \cF_j$ where $O$ is a complex-valued \emph{orthogonal} matrix ($\sum_{j=1}^3 O_{ij} O_{kj} = \delta_{ik}$). For example, a boost in the $x$ direction with rapidity $\rho$ together with a rotation in the $yz$ plane through an angle $\theta$ (a `four-screw' \cite{Synge:1964}) is generated by the transformation matrix
\beq
O = \begin{pmatrix} 1 & 0 & 0 \\ 0 & \cos \gamma & -\sin \gamma \\ 0 & \sin \gamma & \cos \gamma \end{pmatrix}, \quad \quad
\gamma = \theta + i \rho .
\eeq

The complex scalar quantity 
\beq
\Upsilon \equiv -\frac{1}{8} \cF_{ab} \cF^{ab} = \frac{1}{2} \bF \cdot \bF \label{eq:frame-inv}
\eeq 
is frame-invariant. Its real and imaginary parts yield the well-known frame-invariants $\frac{1}{2}(E^2 - B^2)$ and (minus) $\mathbf{E} \cdot \mathbf{B}$, respectively \cite{Dennison:2012vf}. A Faraday field with $\Upsilon = 0$ is called \emph{null}. In the null case, any observer finds that the electric and magnetic fields are orthogonal and of equal magnitude. The superposition of two null fields is not null, in general.

The (observer-dependent) energy density $\mathcal{E} \equiv \frac{1}{2} \left(E^2 + B^2\right)$ and Poynting vector $\mathbf{N} \equiv \mathbf{E} \times \mathbf{B}$ can also be found from the complex three-vector $\bF$, as follows: $\mathcal{E} = \frac{1}{2} \bF \cdot \bF^\ast$ and $\mathbf{N} = \frac{1}{2i} \bF \times \bF^\ast$, with the scalar and vector products extended to complex three-vectors in the straightforward way.

\subsubsection{Maxwell's equations and the vector potential}
The Faraday tensor is governed by (the ``microscopic'' version of) Maxwell's equations,
\beq
\nabla_b F^{ab} = \mu_0 J^a, \quad \quad \nabla_b \Ftil^{ab} = 0,  \label{eq:Maxwell}
\eeq
where $\nabla_a$ denotes the covariant derivative. Here $J^a$ is the four-current density which is necessarily divergence-free ($\nabla_a J^a = 0$). 
The second equation above is equivalent to $\nabla_{[a} F_{bc]} = 0$, known as the Bianchi identity. In the language of forms, $F$ is closed ($dF = 0$ by the Bianchi identity), and thus by Poincar\'e's lemma, $F$ must be locally exact ($F = dA$). Thus, the Faraday tensor can be written in terms of a vector potential $A_a$ as
\beq
F_{ab} \equiv 2 \nabla_{[a} A_{b]} .
\eeq
Due to antisymmetry, it follows that $F_{ab} = 2 \partial_{[a} A_{b]} = \frac{\partial A_b}{\partial x^a} - \frac{\partial A_a}{\partial x^b}$. The Faraday tensor is invariant under gauge transformations of the form $A_{a} \rightarrow A^\prime_a = A_a + \partial_a \chi$, where $\chi$ is any scalar field.

In the absence of charges ($J^a = 0$) we have $\tensor{\cF}{^{ab} _{;b}} = 0 = \cF_{[a b ; c]}$. Then, from a given solution $\cF^{(0)}_{ab}$ one can generate a one-parameter family of solutions $\cF_{ab} = e^{i \varphi} \cF^{(0)}_{ab}$ where $\varphi$ is any complex number. 

\subsubsection{Wave equations\label{sec:waveeqs}}

By taking a derivative of the first equation of (\ref{eq:Maxwell}), re-ordering covariant derivatives, and applying the Bianchi identity, one may obtain a wave equation in the form
\beq
\Box F_{ab} + 2 R_{a c b d} F^{cd} + \tensor{R}{_a ^c} F_{bc} - \tensor{R}{_b ^c} F_{a c} = 2 \mu_0 J_{[a ; b]} ,
\label{eq:waveF}
\eeq
where $R_{a b c d}$ and $R_{a b} \equiv \tensor{R}{^c _{a c b}}$ are the Riemann and Ricci tensors, respectively. In the absence of electromagnetic sources ($J_a = 0$), one may replace $F_{ab}$ with $\cF_{ab}$, if so desired. 

Alternatively, one may derive a wave equation for the vector potential,
\beq
\Box A^a - \tensor{R}{^a _b} A^b - \nabla^a \left( \tensor{A}{^b _{;b}} \right) = -\mu_0 J^a . \label{eq:waveA}
\eeq
The final term on the left-hand side is zero in Lorenz gauge, $\tensor{A}{^a _{;a}} = 0$. 

\subsubsection{Stress-energy tensor}
The stress-energy tensor $T_{ab}$ is given by
\begin{eqnarray}
\mu_0 T_{ab} &\equiv& F_{ac} \tensor{F}{_b ^c} - \frac{1}{4} g_{ab} F_{cd} F^{cd} , \\
 &=& \frac{1}{2} \text{Re} \left( \tensor{\cF}{_a ^c} \cF^\ast_{bc} \right). \label{eq:stress-energy}
\end{eqnarray}
The stress-energy is traceless, $\tensor{T}{^a _a} = \frac{1}{2} \cF_{ab} \cF^{ab\ast} = 0$, and it satisfies the conservation equation $\nabla_b T^{ab} = -\tensor{F}{^a _b} J^b$, which accounts for how energy is passed between the field and the charge distribution.

\subsection{Geometrical optics at leading order\label{subsec:leading}}

Suppose now that the electromagnetic wavelength is short in comparison to all other relevant length scales; and the inverse frequency is short in comparison to other relevant timescales. A standard approach is to introduce a geometrical-optics ansatz for the vector potential $A^a$ into the wave equation (\ref{eq:waveA}) and to adopt Lorenz gauge ($\nabla_a A^a = 0$); see for example Box 5.6 in Ref.~\cite{Poisson:Will:2014}. Another approach \cite{Kristian:1965sz, Ehlers:1967, Anile:1976gq}, which we favour here, is to introduce an ansatz for the Faraday tensor $F_{ab}$ itself. This helps to expedite the stress-energy tensor calculation, and removes any lingering doubts about the gauge invariance of the results obtained. 

We begin by introducing a \emph{geometrical-optics ansatz},
\beq
\cF_{ab} = \cA f_{ab} \exp\left( i \omega \Phi \right) . \label{eq:ansatz}
\eeq 
Here $\omega$ serves as an order-counting parameter; $\Phi(x)$ and $\cA(x)$, the phase and amplitude, respectively, are real fields; and $f_{ab}(x)$, the polarization bivector, is a self-dual bivector field ($f_{ab} = -f_{ba}$, $\widetilde{f}_{ab} = - i f_{ab}$). Loosely, we shall call $\omega$ the `frequency', but with the note of caution that an observer with tangent vector $u^a$ would actually measure a wave frequency of $-\omega u^a \nabla_a \Phi$. 

\subsubsection{Expansion of the wave equation in $\omega$}
Henceforth we shall consider the case of a charge-free region ($J^a = 0$). Inserting (\ref{eq:ansatz}) into the wave equation (\ref{eq:waveF}) (see Sec.~\ref{sec:waveeqs}) yields
\beq
-\omega^2 k^c k_c f_{ab}+  i \omega \left[\left(2 k^c \cA_{;c} + \tensor{k}{^c _{;c}} \cA \right) f_{ab} +  \cA k^c f_{ab;c} \right] + O(\omega^0) = 0,
\eeq
where $k_a \equiv \nabla_a \Phi$. We may proceed by solving order-by-order in $\omega$. 

At $O(\omega^2)$, $k_a k^a = 0$, and thus the gradient of the phase is null. We shall assume henceforth that $k^a$ is future-pointing. It follows inevitably that, as $k^a$ is a gradient and it is null, it must also satisfy the geodesic equation,
\beq
k^b k_{a ; b} = 0.
\eeq
The integral curves of $k^a$ (that is, spacetime paths $x^a(v)$ satisfying $\frac{dx^a}{dv} = k^a$) are null geodesics which lie in the hypersurface of constant phase ($\Phi(x) = \text{constant}$); these are known as the null generators. The null generators may be found from the constrained Hamiltonian
$\mathcal{H}[x^a, k_a] = \frac{1}{2} g^{a b}(x) k_{a} k_{b}$, 
where $\mathcal{H} = 0$ and $k_{a} \equiv g_{a b} \frac{dx^b}{dv}$.

At $O(\omega^1)$, one may split into a pair of transport equations, by making use of the ambiguity in the definitions of $\cA$ and $f_{ab}$ in Eq.~(\ref{eq:ansatz}), viz.,
\begin{eqnarray}
k^a \cA_{;a} &=& - \frac{1}{2} \vartheta \cA,  \label{eq:amplitude} \\
k^c f_{ab;c} &=& 0,  \label{eq:fparallel}
\end{eqnarray}
where $\vartheta \equiv \tensor{k}{^a _{;a}}$ is the expansion scalar. 
Note that (i)  the transport equation for the amplitude $\cA$ ensures the conservation of flux, $\nabla_a \left( \cA^2 k^a \right) = 0$; (ii) by (\ref{eq:fparallel}) the polarization bivector $f_{ab}$ is parallel-propagated along the null generator; and (iii) at leading order the polarization bivector is transverse, $f_{ab} k^b = 0$, which follows from $\tensor{\cF}{^{ab}_{;b}} = 0$ at $O(\omega^1)$.

\subsubsection{Circular polarization}
Conditions (ii) and (iii) are met by the choice 
\beq
f_{ab} = 2 k_{[a} m_{b]}, \label{eq:f0}
\eeq
where $k_a$ is the gradient of the phase, and $m^a$ is any complex null vector satisfying $m_a m^a = m_a k^a = 0$ and $m_a \mbar^a = 1$ (where $\mbar^a$ is the complex conjugate of $m^a$), that is also parallel-propagated along the null generator, $k^b \tensor{m}{^a _{; b}} = 0$. Typically it is constructed from a pair of legs from an orthonormal triad, e.g.~$m^a = \frac{1}{\sqrt{2}} \left( e_1^a + i e_2^a \right)$, and conversely, $e_1^a = \frac{1}{\sqrt{2}} \left( \mbar^a + m^a \right)$ and $e_2^a = \frac{1}{\sqrt{2}}  \left( \mbar^a - m^a \right)$.

The handedness of the circularly-polarized wave depends on the sign of $\omega$ and the handedness of $m_a$. Henceforth, we shall assume that $m_a$ is constructed such that $i \varepsilon_{a b c d} u^a k^b m^c \mbar^d$ is positive for any future-pointing timelike vector $u^a$. The wave is then right-hand polarized (left-hand polarized) if the frequency $\omega$ is positive (negative). 
There remains considerable freedom in the choice of $m_a$, as conditions (ii) \& (iii) and handedness are preserved under the transformation $m_a \rightarrow e^{i \varphi} m_a + \alpha k_a$, where $\varphi$ is any real parameter and $\alpha(x)$ is a real scalar field.

\subsubsection{Stress-energy at leading order}
The circularly-polarized field $\cF_{ab}$ is null at leading order in $\omega$. This can be established by 
inserting Eq.~(\ref{eq:ansatz}) into Eq.~(\ref{eq:frame-inv}) to obtain $\Upsilon = 0$, after noting that $f_{ab} f^{ab} = 0$ for circularly-polarized waves. 

Inserting Eq.~(\ref{eq:ansatz}) into Eq.~(\ref{eq:stress-energy}) gives a leading-order (in $\omega$) result for the stress-energy,
\beq
\mu_0 T_{ab} = \frac{1}{2} \cA^2 k_a k_b .
\eeq
The stress-energy has the form of a null fluid, at this order in $\omega$. It is straightforward to show that $\nabla^b T_{ab} = 0$ by using flux conservation ($\nabla_a \left( \cA^2 k^a \right) = 0$) from property (i) above.
 
 \subsection{Null basis, geodesic deviation and optical scalars\label{subsec:congruence}}
 
 \subsubsection{Null tetrad}
To recap, the leading-order geometrical-optics solution for a circularly-polarized wave is
\beq
\cF_{ab} = 2 k_{[a} m_{b]} \mathcal{A} \exp(i \omega \Phi).  \label{eq:go0}
\eeq
Here $k^a$ is a future-pointing real null vector field ($k_a k^a = 0$) which is the gradient ($k_{[a;b]} = 0$) of the eikonal phase ($k_a = \nabla_a \Phi$), geodesic ($k^b k_{a;b} = 0$) and the null generator of a constant-phase hypersurfaces; and $m^a$ (and its conjugate $\mbar^a$) is a complex null vector field which is unit ($m^a \mbar_a = 1$), right-handed ($i \varepsilon_{abcd} u^a k^b m^c \mbar^d > 0$ for future-pointing timelike $u^a$), parallel-propagated ($k^b m_{a;b} = 0$) and transverse ($m_a k^a = 0$), and thus tangent to constant-phase hypersurfaces ($m^a \nabla_a \Phi = m^a k_a = 0$). 

We may complete the null tetrad by introducing an auxiliary null vector $n^a$ \cite{Poisson:2004}: a future-pointing null vector field satisfying $k_a n^a = -1$ and $m_a n^a = 0$, such that
\beq
\varepsilon^{abcd} = i 4! k^{[a} n^{b} m^{c} \mbar^{d]} .
\eeq
The metric is $g^{ab} = -2k^{(a} n^{b)} + 2 m^{(a} \mbar^{b)}$. 


%

 \subsubsection{Geodesic deviation\label{subsec:geodesic-deviation}}
Consider two neighbouring geodesics (null, spacelike or timelike), $\gamma_0$ and $\gamma_1$, with spacetime paths $x_0^a(v)$ and $x_1^a(v)$ \cite{Poisson:2004} with $v$ an affine parameter. Between $\gamma_0$ and $\gamma_1$, introduce a one-parameter family of null geodesics $x^a(v,s)$, such that $x_0^a(v) = x^a(v,0)$ and $x_1^a(v)=x^a(v,1)$. The vector field $u^a \equiv \partial x^a / \partial v$ is tangent to the geodesics, and thus satisfies $u^b \tensor{u}{^a _{;b}} = 0$. The vector field $\xi^a \equiv \partial x^a / \partial s$ spans the family, though it is not tangent to a geodesic, in general. The identity $\partial \xi^a / \partial v - \partial u^a / \partial s = 0$ (partial derivatives commute) implies that $\xi^a$ is Lie-transported along each geodesic, $\mathcal{L}_u \xi^a \equiv u^b \tensor{\xi}{^a _{;b}} - \xi^b \tensor{u}{^a _{;b}} = 0$. An elementary consequence is that $\frac{d}{dv} \left( \xi^a u_a \right) = 0$, and so $\xi^a u_a$ is constant along each geodesic.  A standard calculation \cite{Poisson:2004} shows that the acceleration of the deviation vector $\xi^a$ is given by
\begin{eqnarray}
\frac{D^2 \xi^a}{dv^2} &\equiv& u^c \left( u^b \tensor{\xi}{^a _{;b}} \right)_{;c} , \nonumber \\
 &=& -\tensor{R}{^a _{bcd}} u^b \xi^c u^d .  \label{eq:geodesic-deviation}
\end{eqnarray}
This is the geodesic deviation equation, which describes how spacetime curvature leads to a relative acceleration between neighbouring geodesics, even if they start out parallel \cite{Poisson:2004}. 

\subsubsection{Optical scalars \& Sachs equations\label{subsec:optical-scalars}}
Now consider the null case with $u^a = k^a$. We may express the deviation vector, restricted to a central null geodesic $\xi^a(v) = \partial x / \partial s |_{s=0}$, in terms of the null basis on that geodesic. Let 
\beq \xi^a = a(v) k^a + b(v) n^a + \overline{c}(v) m^a + c(v) \bar{m}^a ,
\eeq 
where $a$ and $b$ are real and $c$ is complex. After inserting into Eq.~(\ref{eq:geodesic-deviation}) and projecting onto the tetrad, one obtains a hierarchical system of equations:
\begin{subequations}
\begin{eqnarray}
\ddot{b} &=& 0 , \label{eq:bddot} \\
\ddot{a} &=& b R_{nknk} + \overline{c} R_{knkm} + c R_{knk\mbar} , \label{eq:addot} \\
\ddot{c} &=& -b R_{kmkn} - \overline{c} R_{kmkm} - c R_{kmk\mbar} ,  \label{eq:cddot}
\end{eqnarray}
\end{subequations}
where $\ddot{a} \equiv d^2 a / dv^2$, etc., and $R_{k\mbar km} \equiv R_{abcd} k^a \mbar^b k^c m^d$, etc. Note that Eq.~(\ref{eq:bddot}) is consistent with $b = -\xi^a k_a = \text{const.}$, as established above. If one sets $b=0$  then 
\begin{subequations}
\begin{eqnarray}
\ddot{a} &=& \left(\Phi_{00} + \Psi_1 \right) \overline{c} + \left(\overline{\Phi}_{00} + \overline{\Psi}_1 \right) c, \\
\ddot{c} &=& - \Phi_{00} c -  \Psi_0 \overline{c} , 
\end{eqnarray}
\label{eq:cdeviation}
\end{subequations}
where the Ricci and Weyl scalars are given by $\Phi_{00} = \frac{1}{2} R_{kk} = R_{kmk\mbar}$, $\Phi_{01} = \frac{1}{2} R_{km}$, $\Psi_0 = C_{kmkm} = R_{kmkm}$ and $\Psi_1 = C_{knkm}$ (here $C_{kmkm} = C_{abcd} k^a m^b k^c m^d$ and $C_{abcd}$ is the Weyl tensor). 

One may now introduce the ansatz $\dot{c} = \varrho c + \varsigma \overline{c}$, where $\varrho$ and $\varsigma$ are complex functions. From $\mathcal{L}_k \xi^a = 0$ and $b=0$, it follows that $\varrho = m^a k_{a;b} \mbar^b$ and $\varsigma = m^a k_{a;b} m^b$ (see also Appendix \ref{appendix:NP}). Inserting into Eq.~(\ref{eq:cdeviation}) and equating the coefficients of $c$ and $\overline{c}$ leads to a pair of first-order transport equations,
\begin{eqnarray}
\dot{\varrho} &=& - \varrho^2 - \varsigma \overline{\varsigma} - \Phi_{00} , \\
\dot{\varsigma} &=& - \varsigma \left( \varrho + \overline{\varrho} \right) - \Psi_0 .
\end{eqnarray}
These are known as the Sachs equations \cite{Perlick:2004tq}. The real and imaginary parts of $\varrho$ and $\varsigma$ yield the \emph{optical scalars} \cite{Jordan:2013, Kantowski:1968, Frolov:1998wf}: $\varrho = \theta + i \varpi$, $\varsigma = \varsigma_1 + i \varsigma_2$, where $\theta = \frac{1}{2} \tensor{k}{^a _{;a}}$, $\varpi$ and $(\varsigma_1, \varsigma_2)$ are known as the expansion, twist and shear, respectively. The twist is zero for a hypersurface-orthogonal congruence, such as that in the geometrical-optics approximation. 
Kantowski \cite{Kantowski:1968} proved that a (2D) wavefront seen by an observer with tangent vector $u^a$ has principal curvatures $\kappa_{\pm}$ given by $\kappa_{\pm} = (-u^a k_a)^{-1} \left( \theta \pm |\varsigma| \right)$.   

A shortcoming of the Sachs equations is that the optical scalars $\varrho$ and $\varsigma$ necessarily diverge at a conjugate point, where neighbouring rays cross. By contrast, the second-order equation (\ref{eq:cdeviation}) does not suffer from divergences. The optical scalars $\varrho$ and $\varsigma$ can be found from any linearly-independent pair of solutions of Eq.~(\ref{eq:cdeviation}), $c_1$ and $c_2$, by solving
\beq
\begin{pmatrix} \dot{c}_1 \\ \dot{c}_2 \end{pmatrix} = 
\begin{pmatrix} c_1 & \overline{c}_1 \\ c_2 & \overline{c}_2 \end{pmatrix} 
\begin{pmatrix} \varrho \\ \varsigma \end{pmatrix} . \label{eq:optical-c1c2}
\eeq
The inversion breaks down wherever $\text{Im} \left( c_1 \overline{c}_2 \right) = 0$, i.e., at conjugate points. However, note that $c_1$ and $c_2$ are regular at conjugate points, and thus we have a method to find the optical scalars beyond the first conjugate point. 

Similarly, noting that $\dot{a} = -n^b k^a \nabla_a \xi^b = \overline{c} \tau + c \overline{\tau}$, one can find the Newman-Penrose quantity $\tau$ (defined in appendix \ref{appendix:NP}) from a pair of solutions of Eq.~(\ref{eq:cdeviation}) by solving
\beq
\begin{pmatrix} \dot{a}_1 \\ \dot{a}_2 \end{pmatrix} = 
\begin{pmatrix} c_1 & \overline{c}_1 \\ c_2 & \overline{c}_2 \end{pmatrix} 
\begin{pmatrix} \overline{\tau} \\ \tau \end{pmatrix} . \label{eq:optical-c1c2}
\eeq

The complex value $c = \frac{1}{\sqrt{2}} \left(x+iy\right)$ corresponds to a point $(x,y)$ on the wavefront with position vector $\hat{\xi}^a = \overline{c} m^a + c \mbar^a$, with $m^a = \frac{1}{\sqrt{2}} \left( e_1^a + e_2^a \right)$, where $e_i^a$ are orthogonal unit vectors. If $c_1$ and $c_2$ are any pair of linearly-independent solutions of Eq.~(\ref{eq:cdeviation}) then $c(\phi) = \cos \phi \, c_1 + \sin \phi \, c_2$ corresponds to an ellipse in the wavefront. One may show that the principle axes are given by $c_+ = \cos \phi_0 \, c_1 + \sin \phi_0 \, c_2$ and $c_- = -\sin \phi_0 \, c_1 + \cos \phi_0 \, c_2$, where $\tan(2 \phi_0) = 2 \text{Re}(c_1 \overline{c}_2) / \left( |c_1|^2 - |c_2|^2 \right)$, and the semi-major axes $d_\pm = \sqrt{2} |c_\pm|$ are given by $d_+ d_- = 2 \left| \text{Im} ( c_1 \overline{c}_2 ) \right|$ and $d_+^2 + d_-^2 = 2 \left( |c_1|^2 + |c_2|^2 \right)$. It follows that the crosssectional area $A = \pi d_+ d_-$ satisfies the transport equation $\dot{A} = \left(\varrho + \overline{\varrho} \right) A = \vartheta A$. Comparing this with Eq.~(\ref{eq:amplitude}) shows that the square of the wave amplitude, $\mathcal{A}^2$, scales in proportion to the inverse of the crosssectional area of the beam, $A^{-1}$.

\subsection{Differential precession and modified phase\label{subsec:diffprecession}}
In this section we argue that differential precession of the basis $m^a$ along a beam leads to an additional phase term in the leading-order geometrical-optics expansion. The gradient of that phase can be interpreted as a spin-deviation contribution to the tangent vector $k^a$ at order $\omega^{-1}$, whose sign depends on the handedness of the polarization.  

Consider a congruence of null geodesics (see Sec.~\ref{subsec:geodesic-deviation}) with a 2D crosssection seen by an observer with tangent vector $u^a$ and worldline $\gamma$. The crosssection (i.e.~the 2D instantaneous wavefront) is spanned by a basis $m^a = \frac{1}{\sqrt{2}} \left( \hat{e}_1^a + i \hat{e}_2^a \right)$ and $\mbar^a$, such that $k^a m_a = u^a m_a = 0$ and $m^a \mbar_a = 1$. It is natural for an observer to choose a basis that is `straight' in their vicinity, in the sense that $\left. \hat{\xi}^b \nabla_b m^a \right|_\gamma = 0$ for any $\hat{\xi}^a \equiv \overline{c} m^a + c \overline{m}^a$. 
However, a basis that starts out straight does not remain straight, in general, once it is parallel-propagated along the rays in a geodesic null congruence in the presence of a gravitational field. (See e.g.~Ref.~\cite{Nichols:2011pu} for a discussion of differential precession along \emph{timelike} geodesics).

Let $\zeta^a \equiv \left. \xi^b \nabla_b m^a \right|_\gamma$, where $\mathcal{L}_k \xi^a = 0$ and $k^b k_{a;b} = 0$. One may follow steps analogous to those in the derivation of the geodesic deviation equation, Eq.~(\ref{eq:geodesic-deviation}), to derive the \emph{differential precession equation},
\beq
\frac{D \zeta^a}{d v} = -\tensor{R}{^a _{bcd}} m^b \hat{\xi}^c k^d . 
\eeq
Decomposing in the null tetrad, $\zeta^a = \alpha k^a + \dot{c} n^a + \eta m^a$, leads to
\begin{subequations}
\begin{eqnarray}
\dot{\alpha} &=& c \overline{\Psi}_2 , \\
\dot{\eta} &=&  \overline{c} \Psi_1 - c \overline{\Psi}_1.  \label{eq:eta-transport}
\end{eqnarray}
\end{subequations}
in a Ricci-flat spacetime, where $\alpha = \overline{\mu} c + \overline{\lambda} \overline{c}$, $\eta = \overline{c} \chi - c \overline{\chi}$, and $\mu$, $\lambda$ and $\chi$ are Newman-Penrose scalars (see Appendix \ref{appendix:NP}). These scalars can be found from any pair of linearly-independent solutions $(c_1, \alpha_1, \eta_1)$ and $(c_2, \alpha_2, \eta_2)$ satisfying Eqs.~(\ref{eq:cdeviation}) and (\ref{eq:eta-transport}), by inverting 
\beq
\begin{pmatrix} \eta_1 \\ \eta_2 \end{pmatrix} = \begin{pmatrix} \overline{c}_1 & c_1 \\ \overline{c}_2 & c_2 \end{pmatrix} \begin{pmatrix} \chi \\ -\overline{\chi} \end{pmatrix} \quad \text{and} \quad
\begin{pmatrix} \alpha_1 \\ \alpha_2 \end{pmatrix} = \begin{pmatrix} \overline{c}_1 & c_1 \\ \overline{c}_2 & c_2 \end{pmatrix} \begin{pmatrix} \overline{\mu} \\ \overline{\lambda} \end{pmatrix}
\eeq
As for Eq.~(\ref{eq:optical-c1c2}), this procedure fails at a conjugate point.

Suppose that the cross section of the congruence is initially circular and the frame $m^a$ is initially `straight', as shown in Fig.~\ref{fig:differential-precession}(a). After the congruence has passed through a gravitational field, the crosssection will be elliptical, in general; furthermore, the basis will \emph{not} be straight, as shown in Fig.~\ref{fig:differential-precession}(b), due to differential precession. An observer with tangent vector $U^a = \frac{1}{2\beta} k^a + \beta n^a$ (where $\beta > 0$ is a free parameter) will see a wavefront spanned by $m^a$ and $\overline{m}^a$. However, that observer would naturally prefer a `straight' basis $\hat{m}^a = e^{-i \varphi} m^a$, such that $\hat{\xi}^b \nabla_b \hat{m}^a = 0$, where (it is swift to show) the gradient of the phase is
\beq
m^a \nabla_a \varphi = - i \chi .
\eeq

Now consider the leading-order geometric optics solution, Eq.~(\ref{eq:go0}), from the perspective of this observer. With the observer's preference for a locally-straight basis $\hat{m}^a = e^{-i\varphi} m^a$, one could write
\beq
\cF_{ab} = 2 k_{[a} \hat{m}_{b]} \mathcal{A} \exp\left(i \omega \Phi^\prime \right), \quad \quad \Phi^\prime \equiv \Phi + \omega^{-1} \varphi .
\eeq 
The gradient of the \emph{modified phase} $\Phi^\prime$ is
\beq
K_a \equiv \nabla_a \Phi^\prime = k_a + \omega^{-1} \left( i \overline{\chi} m_a - i \chi \mbar_a  + v_a \right) ,
\label{eq:Kfirst}
\eeq
where $v_a m^a = v_a \overline{m}^a = 0$. It is tempting to interpret $K_a$ as an `effective' tangent vector which accounts for the effect of differential precession. Going one step further, we note that one could introduce $M_a \equiv e^{i \varphi} \left(m_a - \omega^{-1} i \chi n_a \right)$ such that  $K^a M_a = 0$, leading to $2 K_{[a} M_{b]} = U_{ab} + i \chi \omega^{-1} W_{ab}$. We shall see in the next section that this argument correctly anticipates part of the geometrical-optics expansion at sub-leading order.

\section{Geometrical optics at higher orders\label{sec:higher-order}}

\subsection{Method\label{subsec:subleading}}

To extend geometrical optics beyond leading order in $\omega$, we shall keep the ansatz (\ref{eq:ansatz}) and expand the self-dual polarization bivector $f_{ab}$ as a power series,
\beq
f_{ab} = f^{(0)}_{ab} + \omega^{-1} f^{(1)}_{ab} + \omega^{-2} f^{(2)}_{ab} + \ldots  \label{eq:fexp}
\eeq
We will expand the self-dual bivectors in the basis $U_{ab}$, $V_{ab}$ and $W_{ab}$ constructed from a twist-free, parallel-propagated null tetrad. The approach is somewhat similar to that in Ref.~\cite{Ehlers:1967}.

Introduce three bivectors (cf.~\cite{Stephani:2003tm}) 
\beq
U_{ab} \equiv 2 k_{[a} m_{b]} , \quad 
V_{ab} \equiv 2 \mbar_{[a} n_{b]} , \quad 
W_{ab} \equiv 2 \left(m_{[a} \mbar_{b]} - k_{[a} n_{b]} \right) ,
\eeq
which are self-dual ($\widetilde{U}_{ab} = -i U_{ab}$, etc.). It is straightforward to verify that (i) the bivectors are parallel-propagated ($k^c U_{ab ; c} = 0$, etc.) and (ii) $U_{ab} V^{ab} = 2$ and $W_{ab} W^{ab} = -4$, with all other inner products zero. Further useful relations are given in Appendix \ref{appendix:Uder}. 

Now let
\beq
f_{ab}^{(j)} = \uuj U_{ab} + \wwj W_{ab}  + \vvj V_{ab} . \label{eq:f1}
\eeq
where $\uuj$, $\wwj$ and $\vvj$ are complex scalar fields, to be determined. At leading order, we choose the circular polarization $f_{ab}^{(0)} = 2 k_{[a} m_{b]} = U_{ab}$ (cf.~Eq.~(\ref{eq:f0})); thus $\uu_0 = 1$ and $\vv_0 = \ww_0 = 0$. 

\subsubsection{Expansion method\label{subsec:subansatz}}

Inserting the ansatz (\ref{eq:fexp}) into the wave equation (\ref{eq:waveF}) yields
\begin{eqnarray}
k^a k_a &=& 0 , \quad k_a \equiv \nabla_a \Phi \quad \Rightarrow \quad k^b \tensor{k}{^a _{;b}} = 0, \\
k^c \nabla_c f_{ab}^{(0)} &=& 0, \quad \quad 
k^c \nabla_c \cA = -\frac{1}{2} \vartheta \cA,  \label{eq:Atransport} \\
k^c \nabla_c f^{(1)}_{ab} &=& \frac{i}{2 \cA} \left[ \Box\left( \cA f_{ab}^{(0)} \right) + 
\cA \left( 2 R_{a c b d} f_{(0)}^{cd} + \tensor{R}{_a ^c} f^{(0)}_{bc} - \tensor{R}{_b ^c} f^{(0)}_{a c} \right)
\right] . \label{eq:f1transport}
\end{eqnarray}
Rather than address the second-order equation (\ref{eq:f1transport}), we may instead expand the equation $\tensor{\cF}{_{ab} ^{;b}} = 0$ order-by-order in $\omega$ to obtain a system of equations 
\beq
f_{ab}^{(j+1)} k^b = \frac{i}{\cA} \nabla^b \left( \cA f_{ab}^{(j)} \right) , \label{eq:f1div}
\eeq
with $f_{ab}^{(0)} k^b = 0$. From Eq.~(\ref{eq:f1}) it follows that the left-hand side of Eq.~(\ref{eq:f1div}) is 
$f_{ab}^{(j+1)} k^b = -\vv_{j+1} \mbar_a + \ww_{j+1} k_a$. Taking projections on the null tetrad,
\beq
i \cA \, \vv_{j+1} = m^a \nabla^b \left(\cA f_{ab}^{(j)} \right) , \quad \quad
i \cA \, \ww_{j+1} = n^a \nabla^b \left(\cA f_{ab}^{(j)} \right) .  \label{eq:vw}
\eeq
Taking further projections of Eq.~(\ref{eq:f1div}) yields
$\mbar^a \nabla^b \left( \cA f_{ab}^{(j)} \right) = 0$ 
and 
$k^a \nabla^b \left( \cA f_{ab}^{(j)} \right) = 0$.
Expanding the former yields a transport equation for $\uu_j$, 
\beq
k^a \nabla_a \uuj = \cA^{-1} \mbar^a \nabla_a \left( \cA \wwj \right) - \lambda \vvj , \label{eq:Du}
\eeq
where $\lambda = \mbar^a n_{a;b} \mbar^b$. 
Expanding the latter yields a transport equation for $\ww_j$ which is consistent with Eq.~(\ref{eq:vw}). 

\subsubsection{Sub-leading order results}
At sub-leading order, inserting $\uu_0 = 1$, $\vv_0 = \ww_0 = 0$ into Eq.~(\ref{eq:vw}) and Eq.~(\ref{eq:w}) yields
\begin{subequations}
\begin{eqnarray}
\vv_{1} &=& i \sigma ,  \label{eq:v} \\ 
\ww_{1} &=& i \cA^{-1} m^a \nabla_a \cA  + i \chi  ,  \label{eq:w} \\
k^a \nabla_a {\uu}_{1} &=& \cA^{-1} \mbar^a \nabla_a \left( \cA \ww_{1} \right) - \lambda \vv_{1} , \label{eq:u1} \\
 &=&  i \left(\cA^{-1} \dbar \delta \cA + \chi \dbar(\ln \cA) + \dbar \chi - \sigma \lambda \right) , \label{eq:u2}
\end{eqnarray}
\end{subequations}
where $\sigma = -m^a k_{a;b} m^b$ and $\chi \equiv \mbar^a m_{a;b} m^b$ are Newman-Penrose quantities \cite{Newman:1961qr} (see Appendix \ref{appendix:NP}). 


\subsubsection{Stress-energy and invariants}
The scalar quantity $\Upsilon \equiv -\frac{1}{8} \cF_{ab} \cF^{ab}$ is given by
\beq
\Upsilon = \frac{1}{2} \cA^2  \left(\omega^{-1} \vv_{1} + \omega^{-2} \left[ \vv_{2} + \uu_{1} \vv_{1} - \ww_{1} \ww_{1} \right] + O(\omega^{-3}) \right) e^{2i\omega\Phi} 
\eeq 
The field is \emph{not} null ($\Upsilon \neq 0$) at sub-leading order if $\vv_{1} = i\sigma \neq 0$. 

Using (\ref{eq:stress-energy}) and (\ref{eq:UVW}), the stress-energy is
\begin{eqnarray}
\mu_0 T_{ab} &=& \frac{1}{2} \cA^2 \left( k_a k_b + 2 \omega^{-1} \text{Re} \left\{ \uu_{1} k_a k_b + \vv_{1} \mbar_a \mbar_b - 2\ww_{1} k_{(a} \mbar_{b)}  \right\} \right. \nonumber \\
&& \; + \omega^{-2} \left[ |\uu_1|^2 k_a k_b + |\vv_1|^2 n_a n_b + 2 |\ww_1|^2 \left(m_{(a} \mbar_{b)} + k_{(a} n_{b)}  \right)  \right.   \nn \\
&&\left. \quad + 2\text{Re} \left(\uu_2 k_a k_b + (\vv_2 + \overline{\uu}_1 \vv_1) \mbar_a \mbar_b - 2(\ww_2 + \overline{\uu}_1 \ww_1) k_{(a} \mbar_{b)} - 2 \overline{\ww}_1 \vv_1 n_{(a} \mbar_{b)} \right) \right] + \nonumber \\
&& \left. \quad \quad + O(\omega^{-3}) \right).  \label{eq:Tsub}
\end{eqnarray}
The subdominant term, at order $\omega^{-1}$, depends on the sign of $\omega$, and thus on the handedness of the circular polarization.

\subsubsection{Parabolic Lorentz transformations and invariants}
Though the direction of $k_a \equiv \nabla_a \Phi$ is fixed, there is residual freedom in the choice of tetrad. Consider a parabolic Lorentz transformation of the form
$k' = k$, $m' = m + B k$, $n' = n + B \mbar + \overline{B} m + B \overline{B} k$,
leading to $U = U'$, $W = W' + 2 \overline{B} U'$, $V = V' + \overline{B} W' + \overline{B}^2 U'$, and thus 
$\vv' = \vv$, $\ww' = \ww + \overline{B} \vv$, $\uu' = \uu + 2\overline{B} \ww + \overline{B}^2 \vv$. Here $B$ is a complex field that is not necessarily parallel-propagated, in general. With the transformation laws $\rho' = \rho$, $\sigma' = \sigma$, $\chi' = \chi + \overline{B} \sigma - B \rho$ and $\lambda' = \lambda + \overline{B}(\overline{\tau} - \overline{\chi}) + \overline{B}^2 \rho + \mbar^a \nabla \overline{B} + \overline{B} k^a \nabla_a \overline{B}$ one may establish that the right-hand sides of Eqs.~(\ref{eq:v}--\ref{eq:u1}) transform in the correct way. Furthermore, one finds
\beq
D \uu^\prime_1 = D \uu_1 + 2 \overline{B} D \ww_1 + \overline{B}^2 D \vv_1 + 2 \ww_1^\prime D \overline{B}
\eeq
where $D = k^a \nabla_a$. 

With the choice $\ww_1^\prime = 0$, that is, $\overline{B} = -\ww_1 / \vv_1$, one has $\uu^\prime_1 = \mathfrak{U}_1$ where
\beq
\mathfrak{U}_1 \equiv \uu_1  - \ww_1^2 / \vv_1 ,  \label{eq:U}
\eeq  
It is straightforward to show that $\mathfrak{U}_1$ is invariant under transformations of the form above, as is 
$\vv_1$. In principle, $\mathfrak{U}_1$ can be calculated via the transport equation
\beq
D \mathfrak{U}_1 = D\uu_1 - 2 \vv_1^{-1} \ww_1 D \ww_1 + \vv_1^{-2}  \ww_1^2 D \vv_1.
\eeq

\subsection{Geometric optics in the Newman-Penrose formalism\label{subsec:NP}}
We now check aspects of the calculation using the Newman-Penrose formulation. 

\subsubsection{Maxwell's equations}
A general self-dual bivector $\cF_{ab}$ can be written as
\beq
  \cF_{ab} = 2 \left( \Phi_{+} U_{ab} + \Phi_{0} W_{ab} + \Phi_{-} V_{ab} \right) ,  \label{eq:Fmaxwell}
\eeq
where $\Phi_{+}, \Phi_0, \Phi_{-}$ are the (complex) Maxwell scalars of spin-weight $+1$, $0$ and $-1$, classified according to their behaviour under rotations of the basis $m^a \rightarrow e^{i\varphi} m^a$. The field equation $\nabla_b \cF^{ab} = 0$ yields four first-order equations (see Appendix \ref{appendix:Uder}):
\begin{subequations}\label{eq:NPmaxwell}
\begin{eqnarray}
D \Phi_0 - \dbar \Phi_- &=& -(\overline{\tau} - \overline{\chi}) \Phi_- + 2 \rho \Phi_0, \label{maxeq1} \\
D \Phi_+ - \dbar \Phi_0 &=& -\lambda \Phi_- + \rho \Phi_+ ,  \label{maxeq2} \\
\Delta \Phi_- - \delta \Phi_0 &=& (2 \gamma - \mu) \Phi_- - 2 \tau \Phi_0 + \sigma \Phi_+ ,  \label{maxeq3} \\
\Delta \Phi_0 - \delta \Phi_+ &=& \nu \Phi_- - 2\mu \Phi_0 + \chi \Phi_+ . \label{maxeq4}
\end{eqnarray}
\end{subequations}
Here $D = k^a \nabla_a$,  $\Delta = n^a \nabla_a$, $\delta = m^a \nabla_a$ and $\dbar \equiv \mbar^a \nabla_a$ are directional derivatives, and the Newman-Penrose coefficients are defined in Appendix \ref{appendix:NP}. These equations were found with the aid of the identities in Appendix \ref{appendix:Uder}. 

We now insert into (\ref{eq:fexp}) a geometrical-optics expansion for the Maxwell scalars that is consistent with Eqs.~(\ref{eq:ansatz}), (\ref{eq:fexp}) and (\ref{eq:f1}), viz.
\begin{subequations}
\begin{eqnarray}
\Phi_+ &=& \frac{1}{2} \cA \left(1 + \omega^{-1} \uu_{1} + \ldots \right) e^{i \omega \Phi} ,  \\
\Phi_0 &=& \frac{1}{2} \cA \left(\phantom{1 + } \; \omega^{-1} \ww_{1} + \ldots  \right) e^{i \omega \Phi} ,  \\
\Phi_- &=& \frac{1}{2} \cA \left(\phantom{1 + } \; \omega^{-1} \vv_{1} + \ldots  \right) e^{i \omega \Phi} ,  
\end{eqnarray}
\end{subequations}
At sub-leading order we deduce 
\beq
\vv_{1} = i \sigma, \quad \quad \ww_{1} = i \delta \left( \ln \cA \right) + i \chi = i \cA^{-1} \mbar_a \delta \left(m^a \cA\right),  \label{eq:vw}
\eeq 
consistent with Eqs.~(\ref{eq:v}) and (\ref{eq:w}), and from Eq.~(\ref{maxeq2}) that
\beq
D \uu_{1} = i \left(\cA^{-1} \dbar \delta \cA + \chi \dbar(\ln \cA) + \dbar \chi - \sigma \lambda \right) ,
\eeq
consistent with Eq.~(\ref{eq:Du}). 

The transport equation for $\uu_{1}$ features second derivatives of the amplitude $\mathcal{A}$ across the wavefront. However, the stress-energy (\ref{eq:Tsub}) at $O(\omega^{-1})$ depends only on the real part of $\uu_{1}$. Isolating the real part,
\beq
D \left( \text{Re}(\uu_{1}) \right) = \frac{i}{2} \left( \cA^{-1} \left(\dbar \delta - \delta \dbar\right) \cA + \chi \dbar(\ln \cA) - \overline{\chi} \delta(\ln \cA)  + \dbar \chi - \delta \overline{\chi} + \overline{\sigma} \overline{\lambda} - \sigma \lambda \right) .
\eeq
and applying the identity
\beq
\dbar \delta - \delta \dbar = (\overline{\mu} - \mu) D - \overline{\chi} \delta + \chi \dbar ,
\eeq
and $D \cA = \rho \cA$ [from Eq.~(\ref{eq:Atransport}) and Appendix \ref{appendix:NP}], leads to
\beq
D\left(\text{Re} (\uu_{1}) \right) =  \frac{i}{2} \left[ \dbar \chi - \delta \overline{\chi} + 2 \chi \dbar(\ln \cA) - 2 \overline{\chi} \delta(\ln \cA) + \rho (\overline{\mu} - \mu) + \overline{\sigma} \overline{\lambda} - \sigma \lambda \right] . \label{eq:Dreu}
\eeq
This transport equation features only first derivatives of the amplitude $\cA$. 

\subsubsection{Transport equations}
The Newman-Penrose quantities $\sigma$, $\rho$, $\chi$, etc., appearing in Eqs.~(\ref{eq:vw}) and (\ref{eq:Dreu}) can (in principle) be found along the null rays using standard transport equations, Eqs.~(\ref{eqs:NPtransport}), once initial conditions are specified. However, Eqs.~(\ref{eq:vw}) and (\ref{eq:Dreu}) also feature the additional quantities $\delta \left( \ln \cA \right)$, $\delta \overline{\chi}$, etc. One can deduce further transport equations by making use of the identity
\begin{eqnarray}
D \delta &=& \delta D - \tau D + \rho \delta + \sigma \dbar , 
\end{eqnarray}
and its complex conjugate. 
Using this, we may obtain a closed system of transport equations for $\delta \ln \cA$, $\dbar \ln \cA$, $\delta \rho$, $\dbar \rho$, $\delta \sigma$, $\dbar \sigma$, $\delta \chi$, and $\dbar \chi$, namely, 
\begin{subequations}
\begin{eqnarray}
D\left(\delta \lnA\right) &=& \delta \rho - \rho \tau + \rho \, \delta \lnA + \sigma \, \dbar \lnA , \\
D\left(\delta \rho \right) &=& 3\rho \delta\rho + \overline{\sigma} \delta \sigma + \sigma \left(\dbar \sigma\right)^\ast - \left(\rho^2 + \sigma \overline{\sigma} \right) \tau + \sigma \dbar \rho , \\
D\left(\delta \sigma \right) &=& 3 \rho \delta \sigma + 2 \sigma \delta \rho - 2 \rho \sigma \tau + \sigma \dbar \sigma - \tau \Psi_0 + \delta \Psi_0 , \\
D\left(\dbar \sigma \right) &=& 3 \rho \dbar \sigma + 2 \sigma \dbar \rho - 2 \rho \sigma \overline{\tau} + \overline{\sigma} \delta \sigma - \overline{\tau} \Psi_0 + \dbar \Psi_0  , \\
D\left(\delta \chi \right) &=& 2 \rho \delta \chi + \chi \delta \rho + \sigma\left(\dbar \chi - (\dbar \chi)^\ast\right) - \overline{\chi} \delta \sigma - \tau \left(\rho \chi - \sigma \overline{\chi} + \Psi_1\right) + \delta \Psi_1 , \\
D\left(\dbar \chi \right) &=& 2 \rho \dbar \chi + \chi \dbar \rho + \left( \overline{\sigma} \delta \chi - \sigma (\delta \chi)^\ast \right) - \overline{\chi} \dbar \sigma - \overline{\tau} \left( \rho \chi - \sigma \overline{\chi} + \Psi_1 \right) + \dbar \Psi_1 .
\end{eqnarray}
\end{subequations}

\subsubsection{Asymptotics}
In a flat (Minkowki) region of spacetime, the transport equations have exact solutions. 
A general pair of solutions to $\ddot{c} = 0$ such that $\rho$ is real are $c_1 = C_1 e^{i\phi_1} (t + \alpha)$ and $c_2 = C_2 e^{i \phi_2} (t + \overline{\alpha})$, where $C_i$, $\phi$ are real constants and $\alpha$ is a complex constant. Without loss of generality for describing the congruence, we choose $C_1 = C_2$ and $e^{i \phi_2} = i e^{i \phi_1} = e^{i (\phi + \pi /4)}$. Solving (\ref{eq:optical-c1c2}) gives
\begin{align}
\rho &= -\frac{1}{2} \left[ (\nu + a + b)^{-1} + (\nu + a - b)^{-1} \right] &&= - \nu^{-1} + a \nu^{-2} - (a^2+b^2) \nu^{-3} + \ldots  \\
\sigma &= \frac{e^{2i\phi}}{2} \left[ (\nu + a + b)^{-1} - (\nu + a - b)^{-1} \right] &&= - e^{2 i \phi} \left(b \nu^{-2} - 2ab \nu^{-3} + \ldots \right) ,
\end{align}
where $a \equiv \text{Re}(\alpha)$ and $b = \text{Im}(\alpha)$. 

By inspection of Eq.~(\ref{eqs:NPtransport}), we can deduce that, in the limit $\nu \rightarrow \infty$, the Newman-Penrose coefficients $\nu$ and $\gamma$ approach constant values; $\cA$, $\rho$, $\chi$, $\tau$, $\lambda$ and $\mu$ decay as $O(\nu^{-1})$; $\sigma$, $\delta \rho$, $\delta \cA$ and $\delta \chi$ decay as $O(\nu^{-2})$; and $\delta \sigma$ decays as $O(\nu^{-3})$. This implies that $\uu_i$, $\ww_i$ and $\vv_i$ scale as $\nu^0$, $\nu^{-1}$ and $\nu^{-2}$, respectively.

\section{Discussion\label{sec:discussion}}

In the previous sections we have extended a geometrical-optics expansion of the Faraday tensor for a circularly-polarized wave through sub-leading order in the expansion parameter $\omega$: see Eqs.~(\ref{eq:ansatz}), (\ref{eq:fexp}), (\ref{eq:f1}),  (\ref{eq:vw}) and (\ref{eq:Dreu}). The method can be extended to higher orders, if required. A key result is the sub-leading order expression for the stress-energy, Eq.~(\ref{eq:Tsub}). This may be re-cast in the following form:
\beq
\mu_0 T_{ab} = \frac{1}{2} \cA^2 K_a K_b + \cA^2 \omega^{-1} i \left( \sigma \overline{m}_a \overline{m}_b - \overline{\sigma} m_a m_b  \right) + O(\omega^{-2}) , \label{eq:Talt}
\eeq
 where
 \beq 
 K_a \equiv k_a + \omega^{-1} \left[ \text{Re}(\uu) k_a -\mathfrak{w} \overline{m}_a - \overline{\mathfrak{w}} m_a \right] + O(\omega^{-2})  , \label{eq:K}
 \eeq
and $\ww = i \chi +  i\cA^{-1}  m^b \nabla_b \cA$. Here $K^a$ is more general than the modified tangent vector in Eq.~(\ref{eq:Kfirst}) of Sec.~\ref{subsec:diffprecession}. Recall that Eq.~(\ref{eq:Kfirst}) was deduced using heuristic arguments about the effect of differential precession on a null congruence; thus it is not surprising to find that Eq.~(\ref{eq:Kfirst}) correctly predicts the differential-precession term $i \chi$ but not the amplitude-gradient term $\cA^{-1} m^b \nabla_b \cA$ in $\ww$, nor the term $\text{Re}(\uu)$ in Eq.~(\ref{eq:K}).

A tentative but appealing interpretation is that the wave's energy propagates principally along $K_a$, rather than $k_a$, and the wave carries with it a transverse stress due to the shear term in (\ref{eq:Talt}). The integral curves of $K^a$ through $O(\omega^{-1})$ are embedded in the constant-eikonal-phase hypersurface. On physical grounds, one may expect $K_a$ to be a null vector, which would imply then that $K_a$ has a component along $n_a$ at $O(\omega^{-2})$, viz.~$- \omega^{-2} \mathfrak{w} \overline{\mathfrak{w}} n_a$. If so, the integral curves of $K^a$ would \emph{not} be embedded in the wavefronts. To investigate this possibility, one could extend the geometrical-optics ansatz (\ref{eq:ansatz}) \& (\ref{eq:fexp}) to next order $O(\omega^{-2})$ following the method herein. 
 
Importantly, the terms at $O(\omega^{-1})$ in Eqs.~(\ref{eq:Talt}), (\ref{eq:K}) depend on the \emph{sign} of $\omega$, and thus on the handedness of the wave (with $\omega > 0$ for right-handed and $\omega < 0$ for left-handed circular polarizations). 
Thus, Eq.~(\ref{eq:Talt}) implies that left- and right-handed wave packets moving through the same spacetime may be deflected in opposite senses, akin to spinning atoms in the Stern-Gerlach experiment. We have identified  a key mechanism that may generate such a splitting: the differential precession across a null congruence that is generated by parallel-propagation through a gravitational field (Fig.~\ref{fig:differential-precession} and Sec.~\ref{subsec:diffprecession}). It is possible that the effect is non-negligible for waves passing close to massive, rapidly-spinning compact objects, such as Kerr black holes. 

In the absence of shear ($\sigma = 0$), the sub-leading order solution is null ($\Upsilon = 0$, see Sec.~\ref{subsec:subansatz}), and we may write the Faraday tensor in the form $\cF_a = 2 K_{[a} M_{b]} \mathcal{A} \exp(i \Phi') + O(\omega^{-2})$ with $K_a$ given by Eq.~(\ref{eq:K}), $M_a \equiv m_a - \omega^{-1} \ww n_a$ and $K_a M^a = 0 + O(\omega^{-2})$; furthermore $\hat{\xi}^a K_a = \hat{\xi}^a \nabla_a \Phi'$ where $\hat{\xi}^a = \overline{c} m^a + c \overline{m}^a$. In short, if $\sigma = 0$ one may write the sub-leading order geometrical-optics solution in an almost-identical form to the leading-order solution (\ref{eq:ansatz}), by modifying the tangent vector $k_a \rightarrow K_a$, the transverse vector $m_a \rightarrow M_a$ and the phase $\Phi \rightarrow \Phi'$. 

One could also extend the investigation of higher-order geometrical optics to other long-range fields with spin; specifically, to neutrinos and gravitational waves. Neutrinos have a definite helicity, and so the differential precession mechanism will split neutrinos from anti-neutrinos. Gravitational waves are typically circularly-polarized with long wavelengths, since they are generated by coherent bulk motions of (e.g.) compact bodies. 

An open question is whether the formulation presented here is of any practical utility in lensing calculations. In other words, can $\uu$, $\vv$ and $\ww$ actually be calculated in practice, via transport equations, for any realistic strong-field lensing scenario? Here there are several practical hurdles, such as (1) finding a parallel-propagated null basis; (2) calculating key quantities such as the Weyl scalars; (3) solving transport equations numerically or otherwise; and (4) handling ray-crossings and conjugate points. 
For the Kerr spacetime, a suitable null basis (1) is known \cite{Marck:1983}, and Weyl scalars (2) can be computed; but (3) finding quantities such as $\dbar \Psi_0$ is challenging, and (4) caustics will arise generically due to axisymmetry. At caustics the Newman-Penrose quantities $\rho$, $\sigma$, etc.~diverge; but it is possible that a second-order formulation, akin to Eq.~(\ref{eq:cdeviation}), can be found to alleviate this issue.

\appendix 

\section{Newman-Penrose formalism\label{appendix:NP}}

The Newman-Penrose (NP) scalars are defined in terms of projections of first derivatives of the null tetrad legs \cite{Newman:1961qr}. For our parallel-propagated basis, three scalars are trivially zero: $\kappa = \pi = \epsilon = 0$. The eight complex scalars used here are defined below:
\begin{subequations}
\begin{align}
\sigma &= -m^a k_{a;b} m^b  ,&
\tau &= - m^a k_{a ; b} n^b , \\
\rho &= -m^a k_{a ; b} \mbar^b , & 
\chi &=  \mbar^a m_{a ; b} m^b , \\
\mu &= \mbar^a n_{a;b} m^b, &
\nu &= \mbar^a n_{a ; b} n^b, \\
\lambda &= \mbar^a n_{a;b} \mbar^b, &
\gamma &= -\frac{1}{2} \left( n^a k_{a ; b} n^b - \mbar^a m_{a ; b} n^b \right) .
\end{align}
\end{subequations}

Certain identities follow from applying $g^{ab} = -k^a n^b - n^a k^b + m^a \mbar^b + \mbar^a m^b$ together with the fact that $k_a$ is a gradient, $k_{[a;b]} = 0$. For example, $\rho$ is purely real due the gradient (twist-free) property of the null tetrad, and $\rho = -\frac{1}{2} \vartheta$ where $\vartheta = \tensor{k}{^a _{;a}}$ is the expansion scalar \cite{Poisson:2004}. Furthermore, $\tau = \beta + \bar{\alpha}$, where $\alpha = \frac{1}{2}\left(k^a n_{a;b} \mbar^b - m^a \mbar_{a;b} \mbar^b \right)$ and $\beta = \frac{1}{2} \left(\mbar^a m_{a;b} m^b - n^a k_{a;b} m^b \right)$. (N.B.~For convenience I have eliminated $\alpha$ and $\beta$ by introducing a new symbol, $\chi \equiv \beta - \overline{\alpha}$). 

The optical scalars of Sec.~(\ref{subsec:optical-scalars}) are simply $\varrho = -\rho$ and $\varsigma = - \sigma$. 

The NP scalars obey a set of transport equations along a null geodesic; see e.g.~Ref.~\cite{Stephani:2003tm}. In a Ricci-flat spacetime ($R_{ab} = 0$), these are 
\begin{subequations} \label{eqs:NPtransport}
\begin{eqnarray}
D\rho &=& \rho^2 + \sigma \overline{\sigma} , \\
D\sigma &=& 2 \rho \sigma + \Psi_0 , \\
D\chi &=& \rho \chi -\sigma \overline{\chi} + \Psi_1 , \\
D\tau &=& \rho \tau + \sigma \overline{\tau} + \Psi_1 , \\
D\lambda &=& \rho \lambda + \overline{\sigma} \mu , \\
D\mu &=& \rho \mu + \sigma \lambda + \Psi_2 , \\
D\nu &=& \overline{\tau} \mu + \tau \lambda + \Psi_3 , \\
D\gamma &=& \tau \overline{\tau} + \frac{1}{2} \left( \overline{\tau} \chi - \tau \overline{\chi} \right) + \Psi_2 .
\end{eqnarray}
\end{subequations}
Here $\Psi_i$ denote the Weyl scalars, defined by 
\begin{align}
\Psi_0 &= C_{kmkm} , &
\Psi_1 &= C_{knkm} , \nn \\
\Psi_2 &= C_{km\mbar n}, &
\Psi_3 &= C_{kn \mbar n}, &
\Psi_4 &= C_{\mbar n \mbar n} ,
\end{align}
where $C_{knkm} \equiv C_{abcd} k^a n^b k^c m^d$, etc., and $C_{abcd}$ is the Weyl tensor. Various identities can be derived using $g^{ac} C_{abcd} = 0$; for example, $\Psi_1 = C_{km\mbar m}$ and $C_{knkn} = C_{m\mbar m\mbar} = \Psi_2 + \overline{\Psi}_2$. 

Some directional derivatives of a parallel-propagated twist-free null basis include
\begin{align}
\delta k^a &= \tau k^a - \rho m^a - \sigma \mbar^a , &
\dbar k^a &= \overline{\tau} k^a - \overline{\sigma} m^a - \rho \mbar^a , \\
\delta m^a &= \overline{\lambda} k^a - \sigma n^a + \chi m^a , &
\dbar m^a &= \overline{\mu} k^a + \rho n^a - \overline{\chi} m^a .
\end{align}
Directional derivatives of the Weyl scalars are given by
\begin{subequations}
\begin{align}
\delta \Psi_0 &= 2\left(\tau + \chi\right) \Psi_0 - 4 \sigma \Psi_1 + \left( \delta C_{abcd} \right) k^a m^b k^c m^d , \\
\dbar \Psi_0 &= 2\left( \overline{\tau} - \overline{\chi} \right) \Psi_0 - 4 \rho \Psi_1 + \left( \dbar C_{abcd} \right) k^a m^b k^c m^d , \\
\delta \Psi_1 &=  \left(\tau + \chi \right) \Psi_1 - 3 \sigma \Psi_2 + \mu \Psi_0 +  \left( \delta C_{abcd} \right) k^a m^b \mbar^c m^d , \\
\dbar \Psi_1 &=  \left(\overline{\tau} - \overline{\chi} \right) \Psi_1 - 3 \rho \Psi_2 + \lambda \Psi_0 +  \left( \dbar C_{abcd} \right) k^a m^b \mbar^c m^d . 
\end{align}
\end{subequations}

Under a change of null basis the Newman-Penrose quantities transform as follows:
\begin{align}
\rho' &= \rho, & \sigma' &= \sigma , \\
\chi' &= \chi + \overline{\alpha} \sigma - \alpha \rho, &
\mu' &= \mu + \overline{\alpha}(\tau + \chi) + \overline{\alpha}^2 \sigma , \\
\tau' &=  \tau + \overline{\alpha} \sigma + \alpha \rho , &
\lambda' &= \lambda + \overline{\alpha}( \overline{\tau} - \overline{\chi} ) + \overline{\alpha}^2 \rho .
\end{align}
and
\begin{equation}
\Psi_0' = \Psi_0, \quad\quad \Psi_1' = \Psi_1 + \overline{\alpha} \Psi_0, \quad\quad \Psi_2' = \Psi_2 + 2\overline{\alpha} \Psi_1 + \overline{\alpha}^2 \Psi_0 .
\end{equation}


\section{Identities for the bivector basis\label{appendix:Uder}}

The following identities are used in calculating the stress-energy tensor,
\begin{subequations}  \label{eq:UVW}
\begin{align}
\tensor{U}{_{a} ^{c}} U^\ast_{bc} &= k_a k_b , &
\tensor{V}{_{a} ^{c}} V^\ast_{bc} &=  n_a n_b , & 
\tensor{W}{_{a} ^{c}} W^\ast_{bc} &= 2m_{(a} \mbar_{b)} + 2k_{(a} n_{b)} , \label{eq:UVW1}
\\
\tensor{V}{_{a} ^{c}} U^\ast_{bc} &= \mbar_a \mbar_b , & 
\tensor{W}{_{a} ^{c}} U^\ast_{bc} &= -2 k_{(a} \mbar_{b)}, &
\tensor{V}{_{a} ^{c}} W^\ast_{bc} &= -2 n_{(a} \mbar_{b)}. \label{eq:UVW2}
\end{align}
\end{subequations}

The Maxwell equations (\ref{eq:NPmaxwell}) are derived by inserting (\ref{eq:Fmaxwell}) into $\tensor{\cF}{_{ab}^{;b}} = 0$ and using the following results:
\begin{subequations}
\begin{eqnarray}
\tensor{U}{_{ab} ^{;b}} &=& \chi k_a + \quad \quad + \rho m_a - \sigma \mbar_a \\
\tensor{V}{_{ab} ^{;b}} &=& \nu k_a + (\overline{\tau} - \overline{\chi}) n_a - \lambda m_a + (\mu - 2\gamma) \overline{m}_a , \\
\tensor{W}{_{ab} ^{;b}} &=& -2 \mu k_a - 2\rho n_a + \quad \quad + 2 \tau \overline{m}_a , \\
U_{ab} \uu^{;b} &=& \delta \uu \, k_a \quad \quad - D\uu \, m_a , \\
V_{ab} \vv^{;b} &=& \quad \quad -\dbar \vv \, n_a \quad \quad + \Delta \vv \, \mbar_a,  \\
W_{ab} \ww^{;b} &=& -\Delta \ww \, k_a + D\ww \, n_a + \dbar \ww \, m_a - \delta \ww \, \mbar_a , \\
\Delta k_a - D n_a &=& (\gamma + \overline{\gamma}) k_a - \overline{\tau} m_a - \tau \overline{m}_a ,\\
\overline{\delta} m_a - \delta \overline{m}_a &=& (\overline{\mu} - \mu) k_a - \overline{\chi} m_a + \chi \mbar_a.
\end{eqnarray}
\end{subequations}

\acknowledgments
With thanks to Luiz Leite, Lu\'is Crispino, Abraham Harte, Antonin Coutant and Jake Shipley for helpful discussions. 
I acknowledge financial support from the Engineering and Physical Sciences Research Council (EPSRC) under Grant No.~EP/M025802/1, and from the Science and Technology Facilities Council (STFC) under Grant No. ST/L000520/1, and from the project H2020-MSCA-RISE-2017 Grant FunFiCO-777740.
\bibliographystyle{apsrev4-1}
\bibliography{go_refs}

\end{document}